\documentclass[prc,amsmath,twocolumn,showkeys,superscriptaddress,preprintnumbers]{revtex4}
\usepackage{gensymb}
\usepackage{graphicx,color}
\usepackage{amssymb}
\usepackage{enumerate}
\usepackage{verbatim}
\usepackage{natbib}
\usepackage{dsfont}
\usepackage{float}
\usepackage[normalem]{ulem}
\usepackage[outdir=./]{epstopdf}

\begin{document}
  \newcommand {\nc} {\newcommand}
  \nc {\Sec} [1] {Sec.~\ref{#1}}
  \nc {\IBL} [1] {\textcolor{black}{#1}} 
  \nc {\IR} [1] {\textcolor{red}{#1}} 
  \nc {\IB} [1] {\textcolor{blue}{#1}} 
  \nc {\IG} [1] {\textcolor{green}{#1}}

%\title{Does polarization better constrain the optical potential than cross section data?}
%\title{Which observable offers better constraints on the optical potential?}
\title{Statistical tools for a better optical model}

\author{M. Catacora-Rios}
\affiliation{National Superconducting Cyclotron Laboratory, Michigan State University, East Lansing, MI 48824}
\affiliation{Department of Physics and Astronomy, Michigan State University, East Lansing, MI 48824-1321}
\author{G.~B.~King}
\affiliation{National Superconducting Cyclotron Laboratory, Michigan State University, East Lansing, MI 48824}
\affiliation{Department of Physics and Astronomy, Michigan State University, East Lansing, MI 48824-1321}
\affiliation{Department of Physics, Washington University, St. Louis, MO 63130, USA}
\author{A.~E.~Lovell} 
\affiliation{Theoretical Division, Los Alamos National Laboratory, Los Alamos, NM 87545}
\author{F.~M.~Nunes}
\email{nunes@nscl.msu.edu}
\affiliation{National Superconducting Cyclotron Laboratory, Michigan State University, East Lansing, MI 48824}
\affiliation{Department of Physics and Astronomy, Michigan State University, East Lansing, MI 48824-1321}

\date{\today}
\preprint{LA-UR-20-28425}

%%%%%%%%%%%%%%%%%%%%%%%%%%%%%%%%%%%%%%%%%%%%%%%%%%%%%%%%%%%%%%%%%%%%%%%%%%%%%%%%%%%%%%%%%%%%%%%%%%%%%%%%%%%%%%%%%%%%%%%%%%%%%%%%%%%

\begin{abstract}
\begin{description}
\item[Background:] Modern statistical tools provide the ability to compare the information content of observables and provide a path to explore which experiments would be most useful to give insight into and constrain theoretical models. 
\item[Purpose:] In this work we study three such tools in the context of nuclear reactions with the goal of constraining the optical potential. 
\item[Method:] The three  statistical tools  examined are:  i)  the principal component analysis; ii) the sensitivity analysis based on derivatives; and iii) the Bayesian evidence. We first apply these tools to a toy-model case, comparing the form of the imaginary part of the optical potential. Then we consider two different reaction observables, elastic angular distributions and polarization data for reactions on $^{48}$Ca at two different beam energies.
\item[Results:]  For the toy-model case, we find significant discrimination power in the sensitivities and the Bayesian evidence, showing clearly that the volume imaginary term is more useful to describe scattering at higher energies. When comparing between elastic cross sections and polarization data using realistic optical models, sensitivity studies indicate that both observables are roughly equally sensitive but the variability of the optical model parameters is strongly angle dependent.  The Bayesian evidence shows some variability between the two observables, but the Bayes factor obtained is not sufficient to discriminate between angular distributions and polarization. 
\item[Conclusions:]  From the cases considered, we conclude that in general elastic scattering angular distributions have similar impact in constraining the optical potential parameters compared to the polarization data. The angular ranges for the optimum experimental constraints can vary significantly with the observable considered.
\end{description}
\end{abstract}


\maketitle

\section{Introduction}
\label{intro}

Nuclear reactions offer an incredibly versatile probe into nuclear structure, nuclear astrophysics, and nuclear applications beneficial to society and are particularly important in the context of rare isotopes \cite{mpla}. One of the most important ingredients in predicting observables for nuclear reactions is the optical potential. The optical potential is an effective interaction between two composite nuclei that incorporates the complexity of the many-body problem into a multi-component multi-parameter complex form. The imaginary part of the optical potential accounts for all the processes that can take flux away from the incident elastic channel into other channels not included in the simplified model. In this work, we will focus on nucleon optical potentials, between either a proton or a neutron and a target nucleus ($U_{NA}$). Because of its effective nature, the nucleon optical potential depends on the mass and charge of the target ($A,Z$), and the energy of the beam ($E_b$). 

While there are many efforts to derive the optical potential starting from the NN interaction (e.g. \cite{rotureau2017,rotureau2018,burrows2018,whitehead2019}), for practical purposes, when computing reactions observables for $A>20$, phenomenological optical potentials are typically used \cite{bg69,ch89,kd2003}. These are obtained from fitting primarily elastic-scattering data, although other observables such as analyzing powers and total cross sections are  sometimes also included in the fitting protocol. 

Over the last few years, there has been substantial progress on quantifying uncertainties in nuclear reactions \cite{lovell2018,king2019,catacora2019}.
We have implemented and applied the Bayesian Markov Chain Monte Carlo method (MCMC) to a variety of reactions \cite{lovell2018}. In \cite{king2019}, we compared the results obtained using the Bayesian approach to the standard $\chi^2$-minimization techniques widely used in the field. Note that $\chi^2$ minimization does not in itself provide any uncertainty but it is common to assume a normal distribution and extract confidence bands from the $\chi^2$ function around the minimum. What we showed in \cite{king2019} is that the uncertainties extracted in this manner were much smaller than those obtained using the Bayesian approach. When we confronted these uncertainties with the data, the empirical coverage demonstrated that for high confidence, the  $\chi^2$-minimization technique largely underestimates the uncertainties, while the Bayesian approach provides an accurate account. These differences could be tracked back to the wrong assumption in the 
$\chi^2$-minimization technique of a Gaussian distribution for the parameters. In addition, when inspecting the correlation between parameters, the $\chi^2$ minimization introduces many types of correlations, while  in the Bayesian approach, only a couple of parameters were found to correlate strongly. As we argue in \cite{king2019}, due to the grid searches in the standard $\chi^2$ minimization, correlations were enhanced as compared to those obtained in the Bayesian MCMC framework.  With these conclusions in mind, we have continued the work on uncertainty quantification using exclusively Bayesian statistics.

In a recent work \cite{catacora2019}, we looked into ways to reduce the uncertainty  from the optical potential by diversifying the data sets or by choosing data with optimum information content. Namely we investigated which parts of the angular distribution matter for constraining the cross sections, and whether data at nearby energies adds further restrictions on the parameters. We have also investigated  the possibility of including polarization data through $iT_{11}$. To diagnose whether including different sets of data mattered, we compared directly the widths of the 95\% confidence intervals.  However comparing directly two confidence intervals  obtained from constraints using two different sets of data has serious ambiguities: what is the appropriate confidence level to make this comparison? and at what angle should the conclusion be drawn? Ultimately we would like to be able to establish without ambiguity which data set (or sets) contain maximum information, resulting in minimum uncertainty. For that, we turn to a wider set of diagnostic tools.

Borrowing from statistical methods applied in many other fields, in this work we inspect three different approaches: i) the principal component analysis that can help to identify whether there is a combination of observables that can simplify the problem; ii) a sensitivity analysis based on derivatives of the the optical potential parameters with respect to observables,  and iii) the Bayesian evidence that integrates the likelihood over the model space to quantify the information content of a given observable. 

These statistical methods are not new; they have been widely applied to other fields and in the last few years have been ported into the nuclear domain. Sensitivity studies  continue to provide insight into which parameters have the largest impact on the observables considered (see \cite{fox2020,ekstrom2019,kejzlar2020} for recent applications to nuclear structure).  
The principal component analyses, coming from the diagonalization of the sensitivity matrix, identify the best combinations of parameters or observables that should be considered to reduce the dimensionality of the model space. It is often used to construct emulators (e.g. \cite{novak2014,sangaline2016,gomez2012}).
Finally, the Bayesian factor, the ratio of the evidence associated with two models, provides an implementation of Occam's razor: the simplest theory compatible with data should be used \cite{trotta2008}. It is also used to discriminate between the information content of different data.  Although calculating the Bayesian evidence is numerically challenging, there are already examples of applications to effective field theory \cite{wesolowski2016} and to nuclear structure \cite{kejzlar2020}.

This paper is organized in the following manner. In section II we introduce the three statistical tools. Section III focuses on a simple toy model to illustrate these same tools. This is followed by the application of these tools to compare the information contained in elastic angular distributions and polarization data (Section IV). 
%We performed calculations for elastic scattering of protons on $^{48}$Ca at 12 MeV and 21 MeV, neutrons on $^{48}$Ca at 12 MeV to investigate which types of data are best suited to constrain the nucleon optical potential in this energy range, using mock data. 
Finally conclusions are drawn in Section V.

%%%%%%%%%%%%%%%%%%%%%%%%%%%%%%%%%%%%%%%%%%%%%%%%%%%%%%%%%%%%%%%%%%%%%%%%%%%%%%%%%%%%%%%%%%%%%%%%%
\section{Statistical methods}
\label{theory}

In this section, we briefly describe three diagnostic tools to quantify the information content of a given set of scattering data. 

\subsection{Principal Component Analysis}

The optical potential  imprints itself on many reaction observables. As such we can consider 
a principal component analysis (PCA) on observable space. Parameter space is randomly sampled $N$ times, 
generating a parameter set $x_i$ for each run used to calculate a set of observables $y_i$.
The parameters and observable data are standardized ({\it i.e}, mean-centered with a standard deviation of 1).
We denote the standardized parameters and observables as $\bar{x}$ and $\bar{y}$, respectively.
The observable covariance matrix, $\mathds{C}$, is then calculated as:
\begin{equation}
\mathds{C}_{ab} = \frac{1}{N-1} \bar{y}_a^T\bar{y}_b \;.
\end{equation}
The principal components of observable space ($e_b$) are the eigenvectors of $\mathds{C}$. The eigenvalues 
$\lambda_b$ corresponding to the eigenvectors $e_b$ can be used to organize the principal components. A larger value 
of $\lambda_b$ means that a larger percentage of the variance in the data is captured by the corresponding $e_b$. 
To visualize the composition of the eigenvectors, $e_b$, we construct the weight matrix,
\begin{equation}
\mathds{W}_{ab} = |\hat{y}_a \cdot e_b| \, ,
\end{equation}
where $\hat{y}_a$ is the $a^{{\rm th}}$ unit vector in the observable space basis. Each column in $\mathds{W}$ is normalized to unity:
\begin{equation}
\sqrt{\sum_i \mathds{W}^2_{ij}} = 1
\end{equation}

We can then assign a weight, $w_a$, to an observable, $y_a$:
\begin{equation}
w_a = \sum_{b} \lambda_b |\mathds{W}_{ab}| \,.
\label{eq:weight}
\end{equation} 
This weight serves as a measure of the importance of the observable across the various principal components and it is often used to identify the most impactful observables.

\subsection{Sensitivity study with derivatives}

Looking at the impact of variations of the observables on the parameters can help determine which optical model parameters are most sensitive to changes in specific observables, and  therefore which observables are most important for constraining the optical model.  A simple approach is to consider directly the covariance $\widetilde{\mathds{C}}_{ia} $ of an observable $\bar{y}_a$ and a parameter $\bar{x}_i$:
\begin{equation}
\mathds{C}_{ia} = \frac{1}{N-1} \bar{x}_i^T\bar{y}_a \;.
\end{equation}

We note that $\mathds{C}_{ia}$ relates to the observable covariance matrix through \cite{sangaline2016}:
\begin{equation} 
\widetilde{\mathds{C}}_{ia} = \sum_b \left\langle \frac{\partial \bar{x}_i}{\partial \bar{y}_b} \right\rangle \mathds{C}_{ba}
\end{equation}
This relation can be inverted to provide the sensitivity:
\begin{equation}
\left\langle \frac{\partial \bar{x}_i}{\partial \bar{y}_a} \right\rangle = \widetilde{\mathds{C}}_{ib} \mathds{C}^{-1}_{ba} \;
\end{equation}
which quantifies the variation in parameter $x_i$ caused by a variation in the observable $y_a$.

\subsection{Bayesian evidence}

As opposed to the frequentist tools based on covariance matrices, Bayesian statistics is based on probability distributions.
For some hypothesis, $H$, data, $d$ and model, $\mathcal{M}$, Bayes' theorem is expressed as:
\begin{equation}
p(H|d,\mathcal{M})= \frac{p(d|H,\mathcal{M})p(H|\mathcal{M})}{p(d|\mathcal{M})}\;,
\label{bayes}
\end{equation}
which states that the posterior probability for the hypothesis, $p(H|d,\mathcal{M})$, taking the data and some assumptions about the model into account, is equal to the likelihood function, $p(d|H,\mathcal{M})$ (the sampling distribution of the data assuming the hypothesis is correct), times the prior probability, $p(H|\mathcal{M})$ (the physical knowledge of the model without any external information) divided by the Bayesian evidence $p(d|\mathcal{M})$. 

While in previous works we have focused on evaluating the posterior distributions of parameters and subsequent confidence intervals of observables, in this work, we focus on the Bayesian evidence. This normalization factor provides a quantification of the information content of a given set of data. For a given model $\mathcal{M}$ with a certain number of parameters $\alpha$, the Bayesian evidence is the integral of the likelihood function times the prior distribution over the entire parameter space:
\begin{equation}
p(d|\mathcal{M})= \int_{\Omega_{\mathcal{M}}} p(d|\alpha,\mathcal{M})p(\alpha|\mathcal{M}) d\alpha_{\mathcal{M}}\;.
\label{integral}
\end{equation}
The explicit calculation of this integral is difficult and computationally demanding as it involves multidimensional integration over all parameters. While the likelihood function might be sharply peaked within the prior range, long tails in the distributions can provide significant contributions to the evidence integral. 

In some applications, the Bayesian evidence integral is approximated analytically (e.g. \cite{wesolowski2016}).  This approximation is valid when the likelihood function is unimodal and the corresponding parameters are Gaussian. However, in our case, these conditions are not met (see for example Fig. \ref{fig-toy-pos9} and \ref{fig-toy-pos65} in Section III). Therefore, to obtain the Bayesian evidence for the optical model, we use the Monte-Carlo approximation to sample full space \cite{kejzlar2019}. Then the integral in Eq. (\ref{integral}) becomes:
\begin{equation}
\hat{p(d|M)}
=\frac{1}{N}\sum_{\alpha_{i}=1}^{N} p(d|\alpha_{i},\mathcal{M})
\label{bayes evidence approx}
\end{equation}
where $N$ parameter sets, $\alpha_{i}$, are drawn randomly from the prior distribution. In practice, when sampling from a prior distribution, it is also important to ensure that the sampled parameters are within the physical region of parameter space. In our application, this means that the optical model parameters should be positive.

%%%%%%%%%%%%%%%%%%%%%%%%%%%%%%%%%%%%%%%%%%%%%%%%%%%%%

\section{Toy-model}
\label{toy-model}

We have conceived a simple case study to illustrate the capabilities of the statistical tools discussed in Section \ref{theory}. 
From decades of phenomenology, it is standard to parameterize the imaginary component of the optical potential with a volume term (Woods-Saxon shape parameterized with $W_v, r_w, a_w$ for the depth, radius and diffuseness) and a surface term (derivative of a Woods-Saxon shape parameterized with $W_s, r_s, a_s$). When using these two components in fits, studies have found that the relative importance of these two components depends strongly on the incident energy of the projectile: at higher energies it is typically  the volume term that dominates. We use this known feature to set up our toy models.

\begin{figure}[t]
\begin{center}
\includegraphics[width=0.23\textwidth]{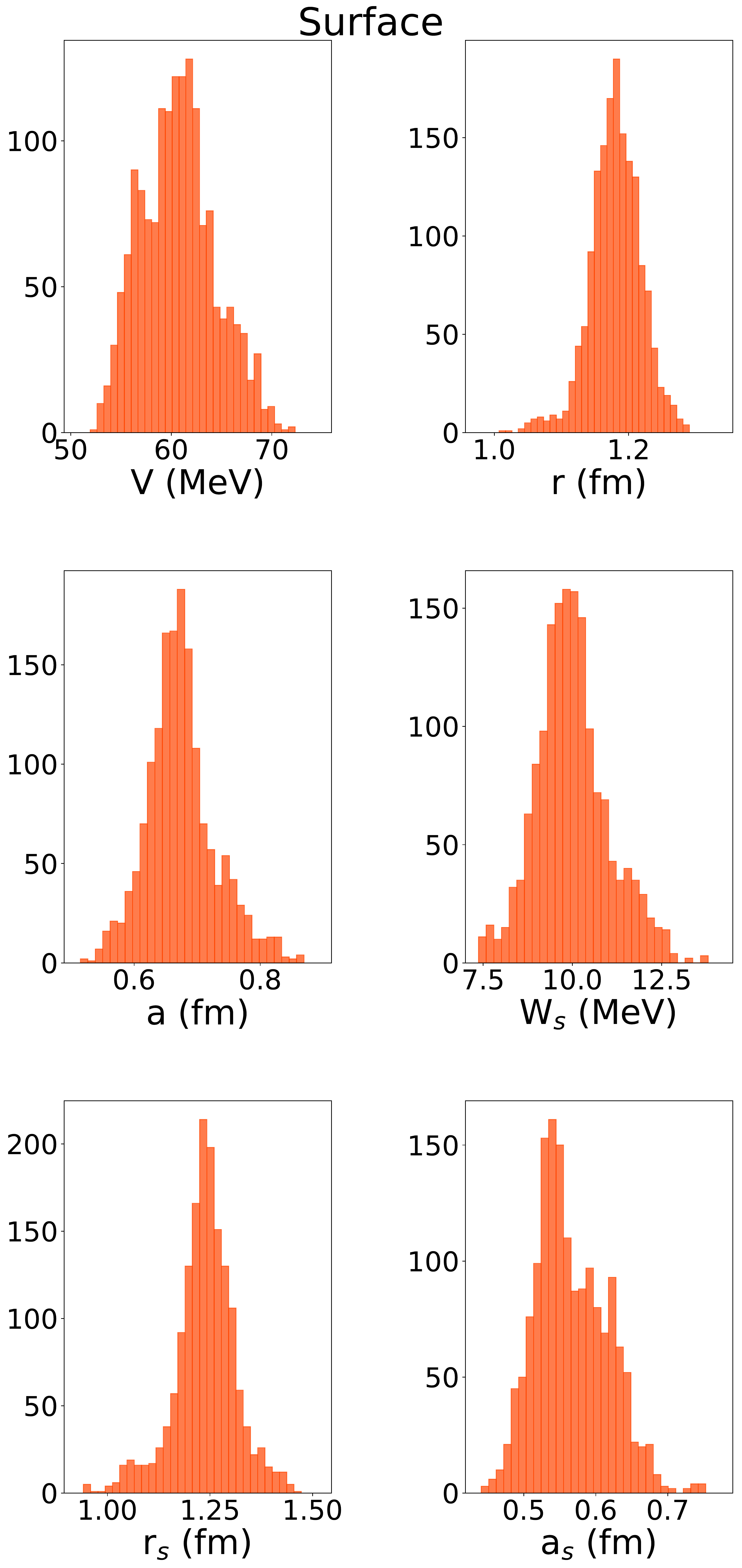}
\includegraphics[width=0.23\textwidth]{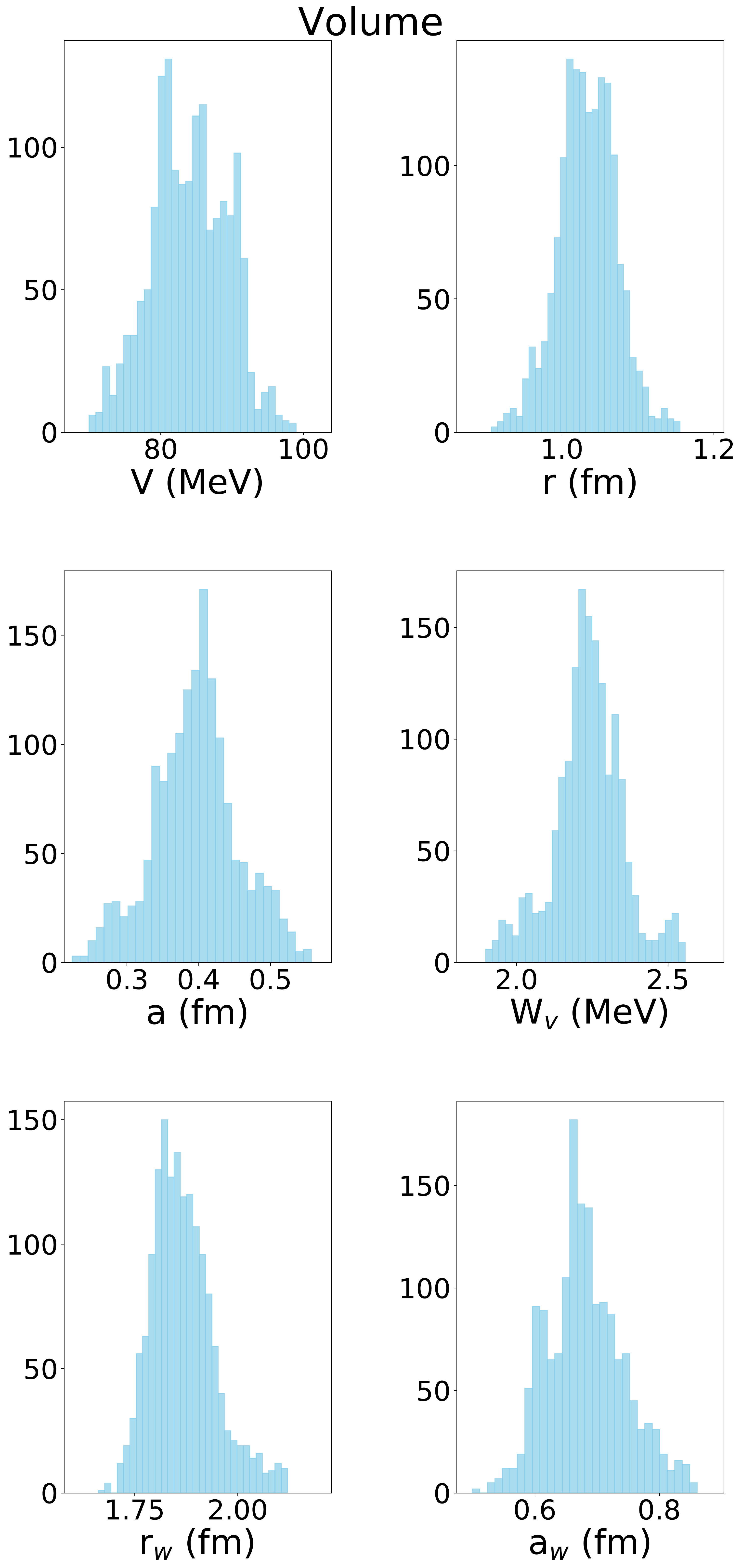}
\end{center}
\caption{$^{48}$Ca(p,p) at 9 MeV parameter posterior distributions: surface model (orange) and volume model (blue).}
\label{fig-toy-pos9}
\end{figure}
\begin{figure}[h!]
\begin{center}
\includegraphics[width=0.23\textwidth]{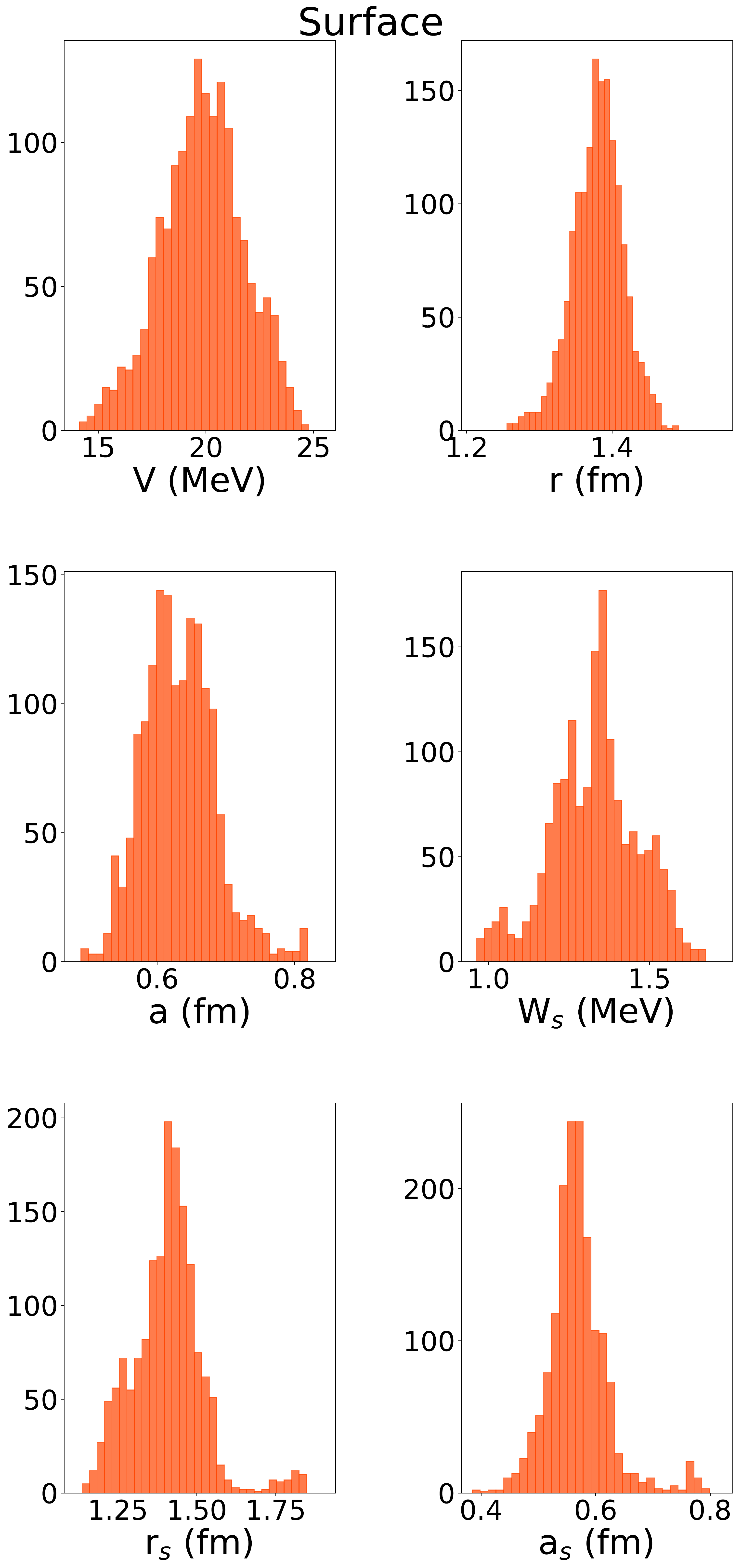}
\includegraphics[width=0.23\textwidth]{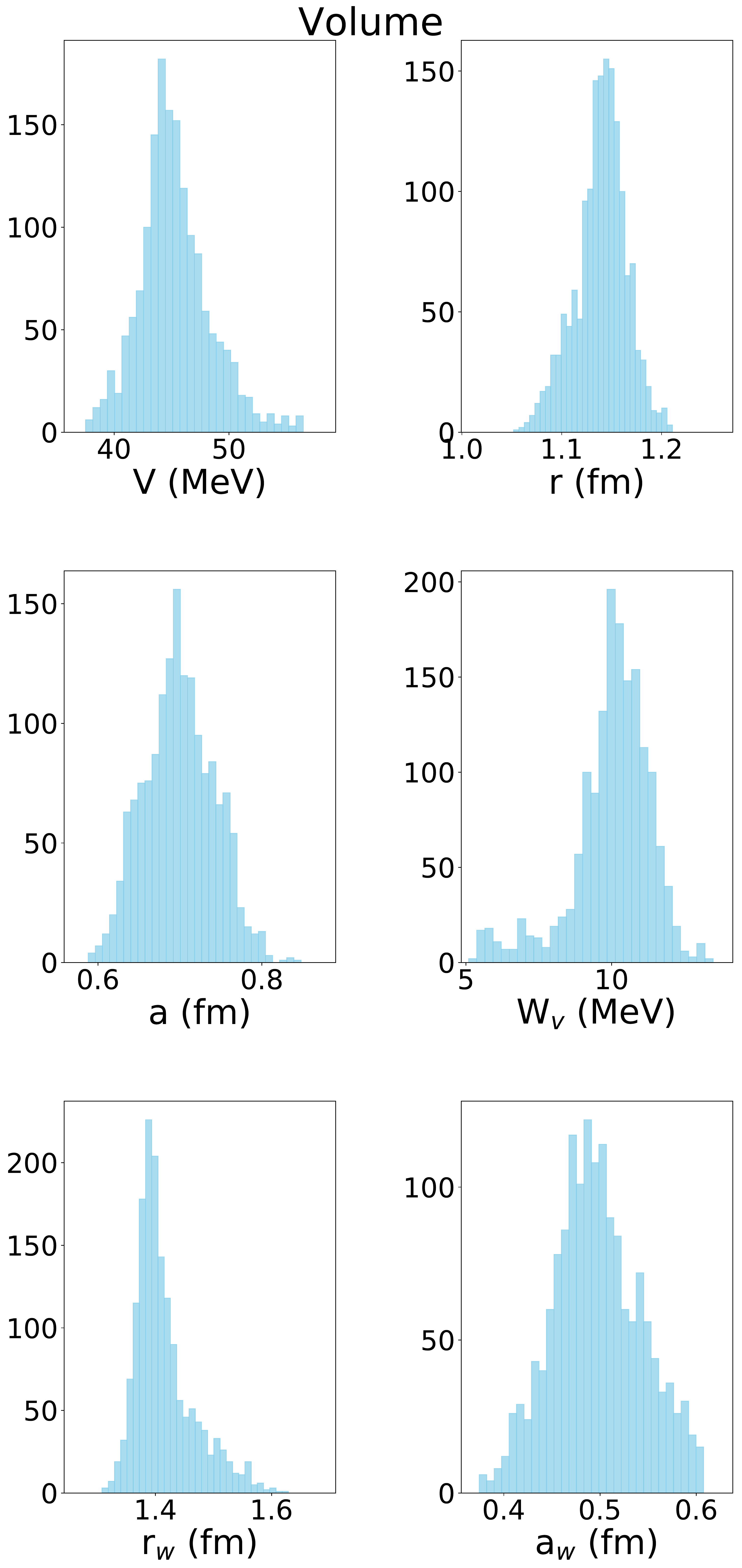}
\end{center}
\caption{$^{48}$Ca(p,p) at 65 MeV parameter posterior distributions: surface model (orange) and volume model (blue).}
\label{fig-toy-pos65}
\end{figure}
We  consider proton elastic scattering on $^{48}$Ca at $E_{lab}=9$ MeV and $65$ MeV within two extreme optical-models: the {\it surface} model that includes only the surface term $W_s$ (while setting $W_v=0$) and the {\it volume} model that includes only the volume term $W_v$ (while setting $W_s=0$).
We use mock data generated from the global optical potential \cite{kd2003} at the appropriate energy, and include a 10\% error, similarly to what was done in \cite{catacora2019}.
For calibration, we perform the MCMC simulation with $1600$ pulls, allowing in each case six parameters to vary. The parameters for the surface model are $V,r,a$ for the real part and $W_s,r_s,a_s$ for the imaginary, and the parameters for the volume model are $V,r,a$ for the real part and $W_v,r_w,a_w$ for the imaginary term. 
The parameters for the spin-orbit term and the Coulomb force were kept fixed and set to the values in \cite{kd2003}.
In addition, wide Gaussian priors, centered at the values of \cite{bg69}, are used as in previous work \cite{catacora2019}, which ensures that the process is data driven. 
Calculations are performed with the suite of codes QUILT-R \cite{quilt-r} (more details on the MCMC calculations can be found in \cite{lovell2018}).  

Figs. \ref{fig-toy-pos9} and \ref{fig-toy-pos65}  display the parameter posterior distributions for the surface model (orange) and the volume model (blue) for $^{48}$Ca at $E_{lab}=9$ MeV and $65$ MeV, respectively. 
It is immediately obvious that changing the shape of the imaginary term affects the real part of the interaction: while the mean of the priors for $V,r,a$ are the same in both calculations, there are large shifts in the peaks of the posterior distributions for these parameters. This is nothing new to the field: the system finds a different minimum depending on the shape of the imaginary term. The  posteriors of the imaginary parameters can also have significantly different widths. For example for 65 MeV, both $W_s$ and $r_s$ have very broad distributions compared to their counterparts $W_v$ and $r_w$, while the distribution for $a_s$ is narrower than that for $a_w$. 
From these distributions alone it is not possible to decide which model would be more appropriate to describe each reaction.  
%Typically, one would look at the $\chi^2$ distribution:  the peak of the distribution obtained with the volume term is shifted to lower values when compared to those obtained using the surface imaginary term. Here we argue that a lot more can be learned by going beyond the $\chi^2$ and proceeding with a full statistical analysis of these results.

\begin{figure*}[t]
\begin{center}
\includegraphics[width=0.35\textwidth]{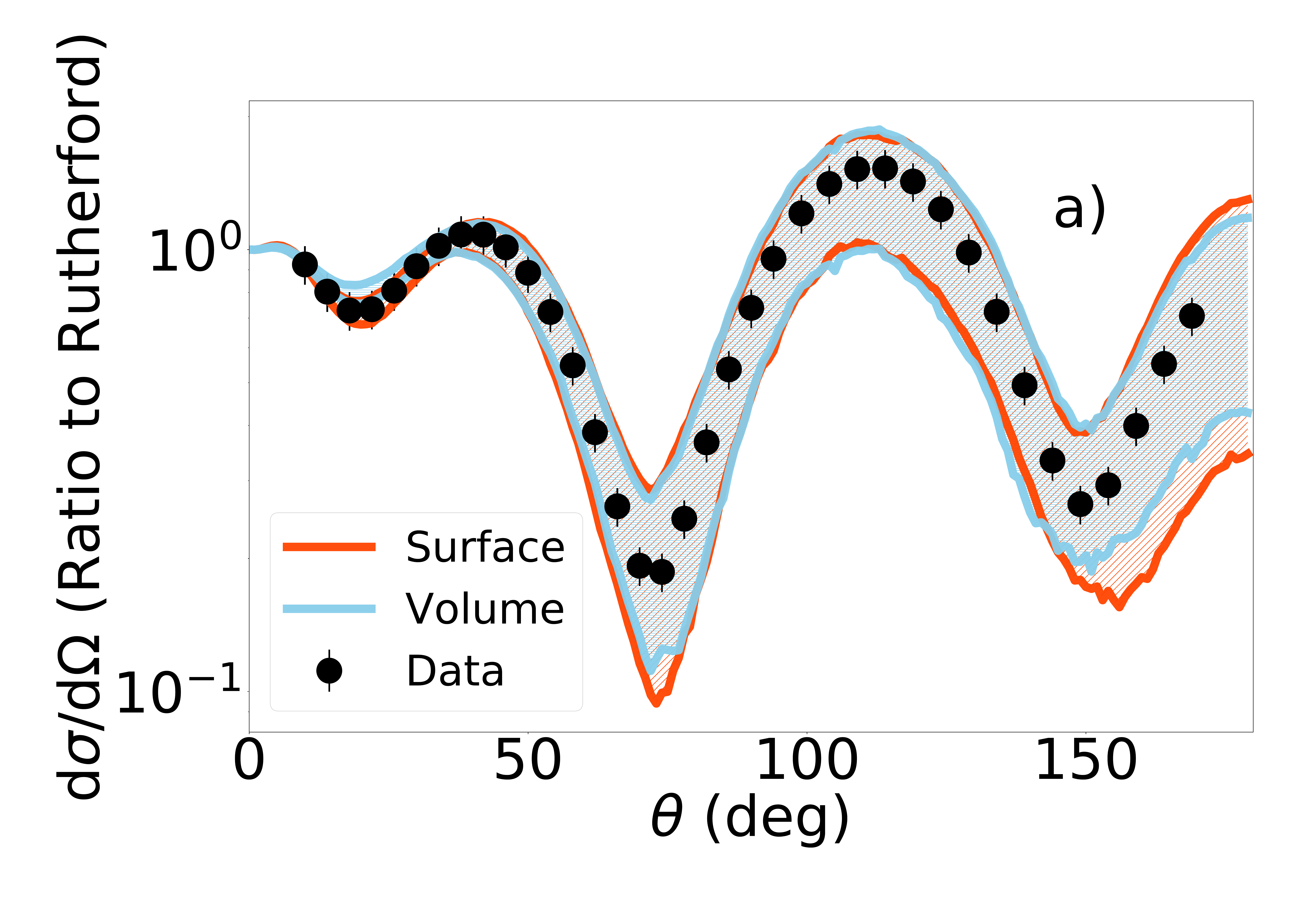}
\includegraphics[width=0.35\textwidth]{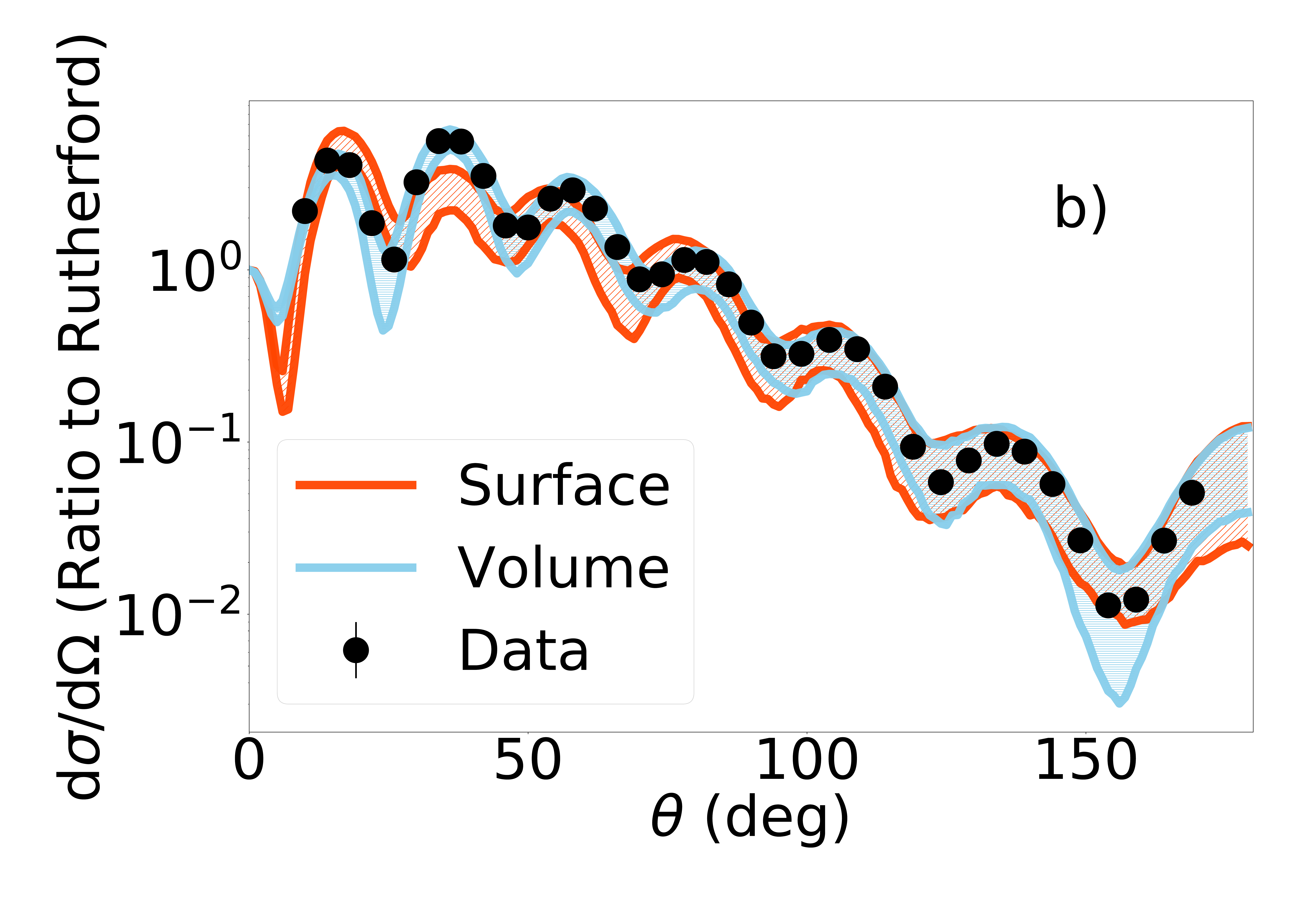}
\end{center}
\caption{Angular distribution 95\% confidence intervals for $\frac{d\sigma}{d\Omega}$ for $^{48}$Ca(p,p) at (a) 9 MeV and (b) 65 MeV: surface model (orange) and volume model (blue). Mock data generated with \cite{kd2003} (black circles).}
\label{fig-toy-ci}
\end{figure*}

One might next consider the observable itself. Fig.\ref{fig-toy-ci} a) and b) contain the 95\% confidence intervals for the angular distributions  of $^{48}$Ca(p,p) at $9$ MeV and $65$ MeV, respectively, using either the surface model (orange) or the volume model (blue). For the 9 MeV reaction, the confidence intervals for the surface and the volume model are roughly the same throughout the angular range. A close inspection of the widths of the confidence intervals at 65 MeV show that at forward angles (within the first couple of diffraction peaks) the volume model provides a narrower uncertainty compared to the surface one, while at backward angles the confidence intervals for the volume model are slightly wider than those generated with the surface model. Still, we are left not knowing which model is best to describe the data.
\begin{figure}[h!]
\begin{center}
\includegraphics[width=0.23\textwidth]{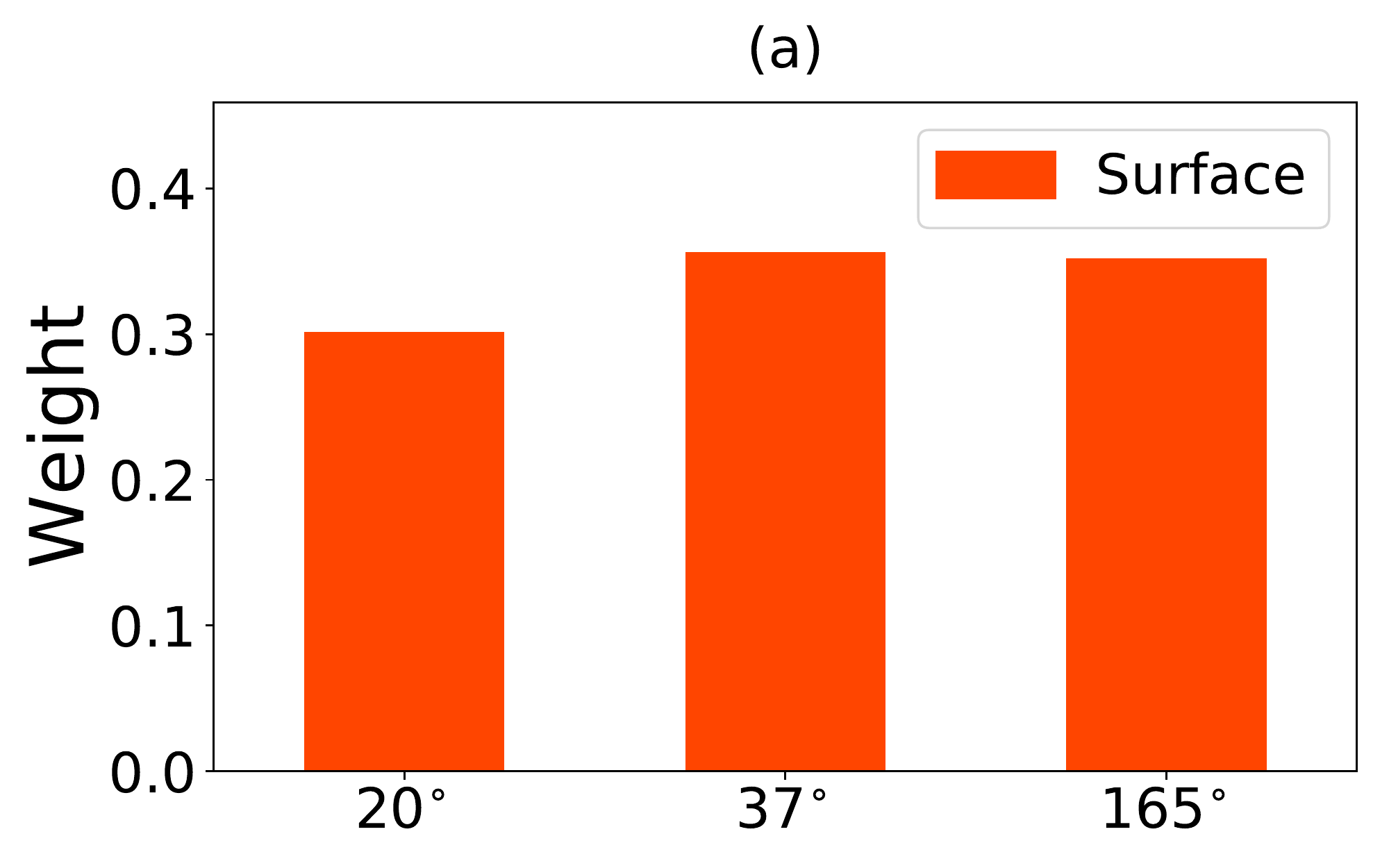}
\includegraphics[width=0.23\textwidth]{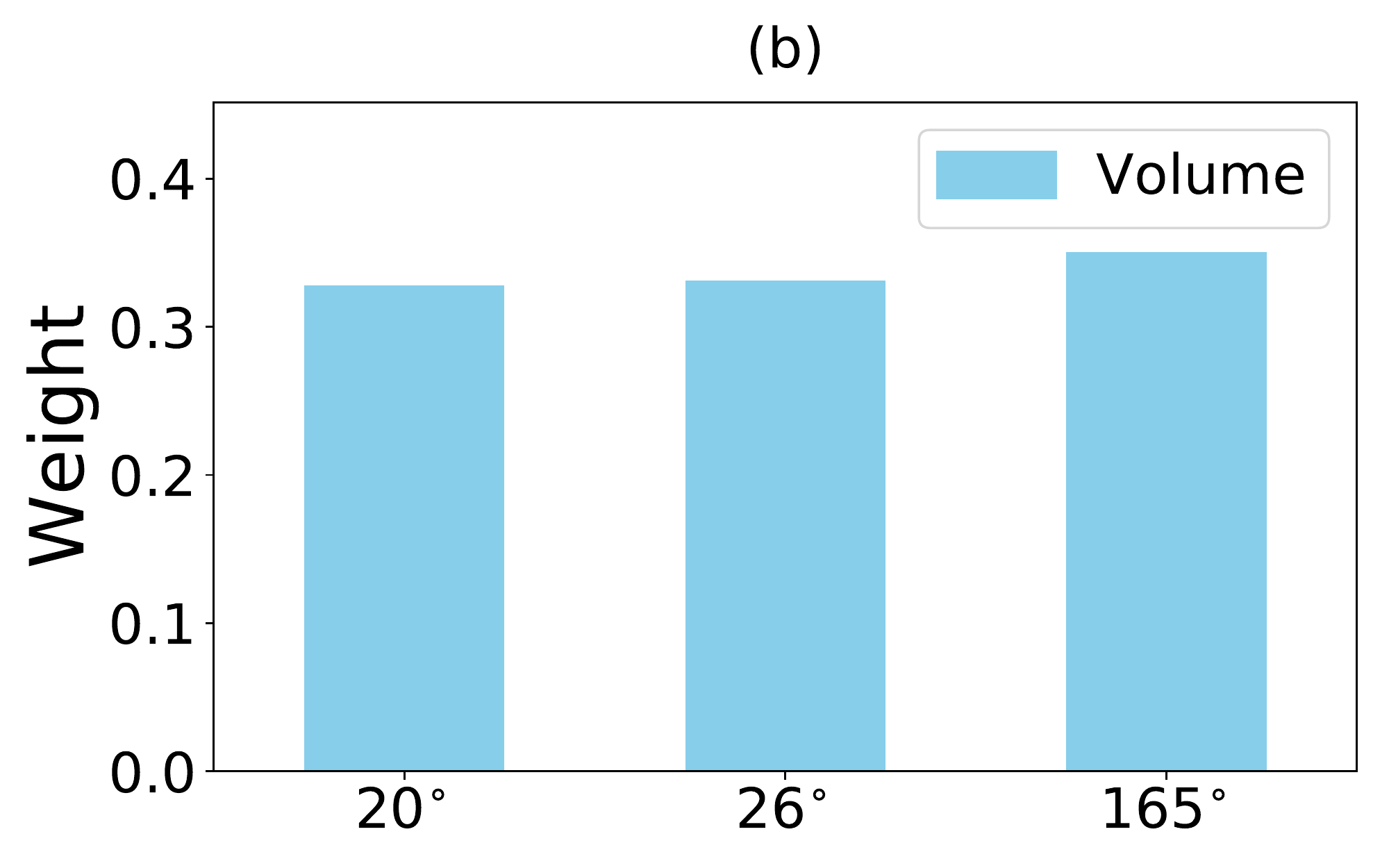}
\end{center}
\caption{Principal component analysis for $^{48}$Ca(p,p) at 65 MeV using $\frac{d\sigma}{d\Omega}$ data. Shown are the weights $w_a$ obtained for three different scattering angles: a) for the surface model and b) for the volume model.}
\label{fig-toy-pca}
\end{figure}

We next consider the weights $w_a$ generated from the principal component analysis as described in Sec. \ref{theory}. 
Pulling from the posterior distributions, we construct the observable  $\frac{d \sigma}{d \Omega}$ covariance matrix for specific angles, which is subsequently diagonalized to obtain the principal components of observable space and the respective weights as described in Section \ref{theory}. Plotted in Fig. \ref{fig-toy-pca} are the  weights corresponding to the surface model (orange) and the volume model (blue), for $^{48}$Ca(p,p) at 65 MeV, at three different angles corresponding to the forward direction,  the first peak of the angular distribution, and the backward direction.
In both models, all components are of roughly equal weight and no  principal component pops out in the analysis.  The same result is obtained if  a fine discretization is included over the whole angular range. For all our applications for the elastic angular distributions and other reaction observables, the weights resulting from the observable PCA are approximate equal. This makes PCA less useful in reducing the dimensionality of the observable space. For this reason we do not include PCA in the applications discussion in Section IV.

\begin{figure}[t]
\begin{center}
\includegraphics[width=0.49\textwidth]{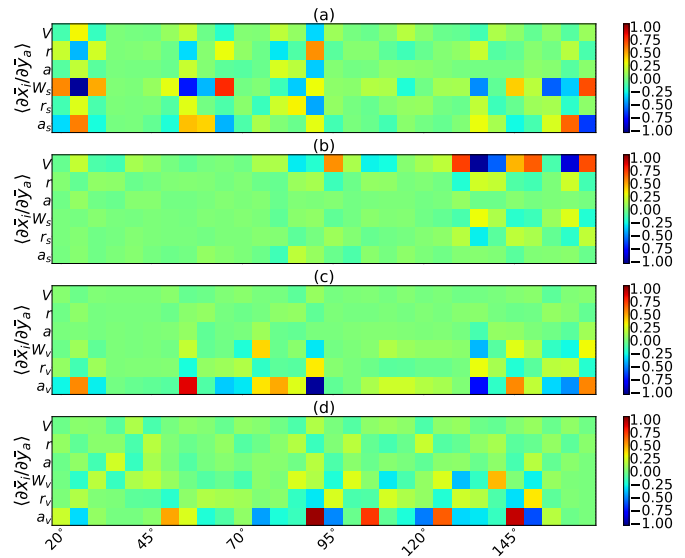}
\end{center}
\caption{Parameter sensitivities for $^{48}$Ca(p,p): a) surface model at 9 MeV, b) surface model at 65 MeV, c) volume model at 9 MeV and d) volume model at 65 MeV.}
\label{fig-toy-sen}
\end{figure}

\begin{table}[b]
\centering
\begin{tabular}{|c | r | r | r |  }
\hline
\text{Energy } & $9$ MeV  & $65$ MeV \\ 
\hline
\text{ Evidence(surface) } & 1.06 &  0.02 \\ 
\text{ Evidence(volume) } &  0.65 &  0.13 \\ 
\text{ Bayes Factor }		&     0.6     &     6.9 \\    
\hline
\end{tabular}
\caption{Bayesian evidence (multiplied by $10^{-3}$) for the surface model (2nd row) and the volume model (3rd row) for both beam energies considered (1st row). The ratio between the Bayesian evidence of the volume model over that with the surface model is in the 4th row (the Bayes' factor).}
\label{tab:toy}
\end{table}

More interesting are the sensitivities $\left\langle \frac{\partial \bar{x}_i}{\partial \bar{y}_a} \right\rangle $ introduced in Section \ref{theory}.
As for the PCAs, the sets of parameters are drawn from the posterior distributions (Figs. \ref{fig-toy-pos9} and \ref{fig-toy-pos65}) and the respective covariance matrices are then computed as described in Section \ref{theory} .
The average, $\langle \frac{d \bar x_i}{d \bar y_a} \rangle$, is obtained for  the elastic angular distributions for angles in the range $\theta=20 -165 ^\circ$ in intervals of $5^\circ$. 

Fig. \ref{fig-toy-sen} contains the results for these sensitivities using the surface model in panels a) and b), and using the volume model in panels c) and d). Along the x-axis are the various angular bins and along the y-axis are the optical potential parameters considered in the Bayesian MCMC. The darker reds and darker blues correspond to large positive and large negative sensitivities, respectively. Results for the reaction at $9$ MeV  show that in the surface model, it is mostly $W_s$ that is constrained by the data, although there is still significant sensitivity to $a_s$. At higher energies the imaginary term is no longer constrained in the surface model, only the real depth becomes sensitive. On the contrary, for the volume model at 9 MeV, the angular distribution using the volume model is not able to constrain the imaginary depth and is mostly sensitive to the imaginary diffuseness.  Depending on the model, one might also make different choices for which angles to measure in the angular distribution.

\begin{table*}[t!]
\centering
\begin{tabular}{c | c | c | c | c | c | c | c }
\hline \hline 
\textbf{Observable} & \textbf{E} (MeV) & \textbf{V} (MeV) & \textbf{r} (fm) & \textbf{a} (fm) & \textbf{W$_\mathrm{s}$} (MeV) & \textbf{r$_\mathrm{s}$} (fm) & \textbf{a$_\mathrm{s}$ (fm) }  \\ \hline
\textbf{$\frac{d\sigma}{d\Omega}$} & 12.0 & 59.48 (4.12) & 1.173 (0.052) & 0.699 (0.051) & 9.476 (0.960) & 1.294 (0.084) & 0.571 (0.049) \\ 
\textbf{$iT_{11}$} & 				12.0 & 60.65 (5.22) & 1.159 (0.057) & 0.699 (0.067) & 9.704 (0.954) &  1.273 (0.079) & 0.595 (0.080)  \\ \hline 
\textbf{$\frac{d\sigma}{d\Omega}$} & 21.0 &  55.57 (4.11) &1.178 (0.052) & 0.661 (0.057) & 7.857 (0.767) & 1.297 (0.083) & 0.572 (0.051)  \\ 
\textbf{$iT_{11}$} & 				21.0 & 57.16 (4.44) & 1.165 (0.047) & 0.691 (0.046) & 8.011 (1.007) & 1.260 (0.073) & 0.579 (0.076)  \\ \hline \hline
\end{tabular}
\caption{Characteristics of the posteriors for the two $^{48}$Ca(p,p) elastic observables considered; the second column is the beam energy; the next three columns provide the means (standard deviations) for the depth, radius and diffuseness of the real part of the optical potential; the last three columns provide the means (standard deviations) for the depth, radius and diffuseness of the imaginary surface terms of the optical potential. }
\label{tab:parameters}
\end{table*}

Our final tool in the tool-set is the Bayesian evidence and Bayes factor for model selection \cite{wiki}. As defined in Section \ref{theory}, the Bayesian evidence provides a direct measure of the information content of a given model in light of a set of data. It serves as a tool to compare different models.
As mentioned earlier, our problem is not amenable to an analytic treatment of the evidence integral thus it is very important to collect enough statistics to ensure convergence of the integral. We have studied the convergence of the evidence with the number of pulls and find that very large statistics need to be collected. The numbers presented in Table \ref{tab:toy} correspond to $1,500,000$  pulls from a  Gaussian distribution 20\% wide, that still ensures that the parameters do not become negative, so they can be restricted to the physical region.
The Bayesian evidence for the surface model is larger than that for the volume model for the reaction at $9$ MeV, as one might expect.
In contrast, the Bayesian evidence for the volume model is larger than that for the surface model at $65$ MeV and clearly indicates that the volume model will contain more information than the surface model at this energy.  

Overall, we find varying success between the statistical methods investigated in this toy problem.  The differences in the Bayesian parameter posterior distributions and observable confidence intervals along with the principal component analysis do little to discriminate between the surface and volume models considered.  On the other hand, we see that the parameter sensitivities and the Bayesian evidence have significant power in discriminating between the two models. These tools will now be applied to realistic cases.

%%%%%%%%%%%%%%%%%%%%%%%%%%%%%%%%%%%%%%%%%%%%%%%%%%%%%%%%%%

\section{Comparing elastic angular distributions with polarization data}
\label{results}

We now use the tools introduced above to explore the information content of two types of reaction observables. In this field it is generally easier to measure the elastic angular distributions than polarization data. However one might ask whether polarization data are best to constrain the optical model. In this context,  we have studied $^{48}$Ca(n,n) at 12 MeV and $^{48}$Ca(p,p) at 12 MeV and 21 MeV, and illustrate  in this section the usefulness of using sensitivities and evidence in the analysis of these reactions. 
%We also studied $^{208}$Pb(n,n) at 30 MeV;  $^{208}$Pb(p,p) at 30 MeV and 61 MeV, but instead of showing all the details we only make remarks on the corresponding sensitivities at the end of Section IV.B and on the corresponding Bayesian evidences at the end of Section IV.C. (see \cite{catacora2019} for other applications to these reactions).
%A selection of these are shown in this section to illustrate the benefits and limitations of these various diagnostic tools.

As before, in this section we use mock data, generated from the global optical potential \cite{kd2003} including a 10\% error, following \cite{catacora2019} (see Section \ref{realdata} for a comparison with real data). For the polarization data, we introduced a lower limit for the error, determined by 5\% of the maximum $iT_{11}$ value.
We initialized the optical potential parameters using the global optical model \cite{bg69}, and use wide Gaussian priors as before.

Next, we inspect the parameter posterior distributions. Instead of showing the full posterior distributions we summarize in Table I the means and standard deviations for the six parameters when considering $^{48}$Ca(p,p) at 12 MeV and 21 MeV. We show the results when using either $\frac{d \sigma}{d \Omega}$ data or $iT_{11}$ data in the optimization procedure. 
%For completeness, the full posterior distributions for each of the six parameters are shown in the  panels a-f of Fig. \ref{fig-post12} and Fig. \ref{fig-post21}, together with the resulting $\chi^2$ and likelihood distributions (panels g and h).  The results using $\frac{d \sigma}{d \Omega}$ data are shown in red (dark) and the results using $iT_{11}$ data are shown in blue (light).
We find that the distributions are mostly overlapping, indicating that the angular distributions and the polarization data lead to very similar minima in parameter space. Some posteriors  present a semi-bimodal structure or extended tails.  This is reflected in slight differences in the means and standard deviations shown in Table \ref{tab:parameters}. Ultimately both sets of data lead to an identical likelihood function and therefore similar goodness of fit.

\subsection{Confidence intervals resulting from the fit}
\label{confband}

\begin{figure*}[t]
\begin{center}
\includegraphics[width=0.75\textwidth]{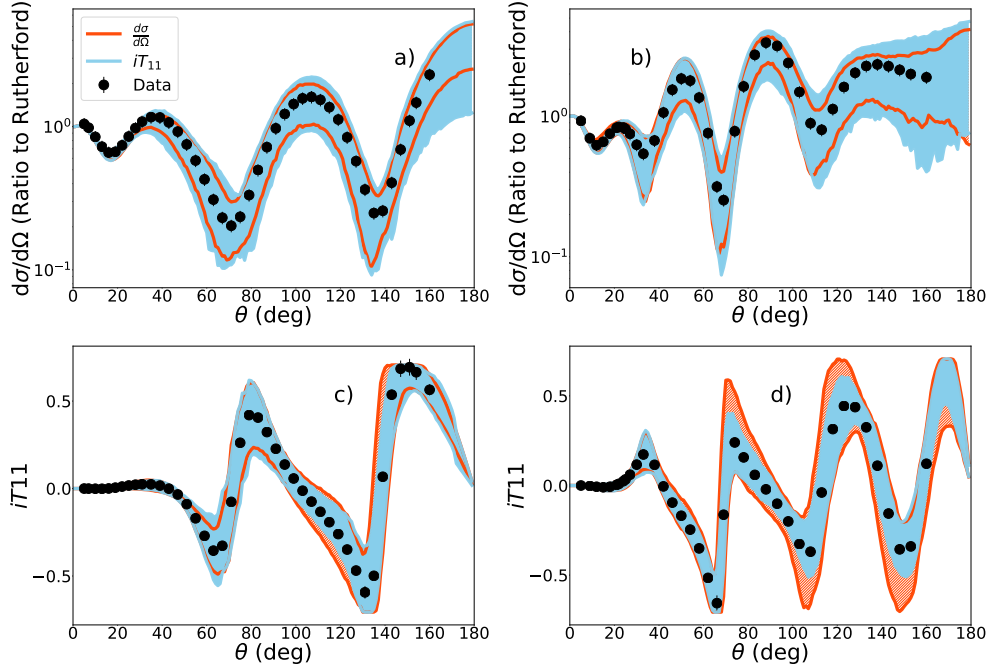}
\end{center}
\caption{95\% confidence intervals calculated using $\frac{d\sigma}{d\Omega}$ data (red bands)  or  $iT_{11}$  data (blue bands) in the optimization procedure: a) $d\sigma/d\Omega$for $^{48}$Ca(p,p) at 12 MeV;  b) $d\sigma/d\Omega$ for $^{48}$Ca(p,p) at 21 MeV; c) $iT_{11}$ for $^{48}$Ca(p,p) at 12 MeV;  d) $iT_{11}$ for $^{48}$Ca(p,p) at 21 MeV.}
\label{fig-ci}
\end{figure*}

In Fig. \ref{fig-ci}, we show  the 95\% confidence intervals for elastic angular distributions (top) and the polarization distributions (bottom) obtained when either the elastic angular distribution data are used  (orange interval) or the polarization data are used (blue interval). 
Figs. \ref{fig-ci}(a) and (c) correspond to $^{48}$Ca(p,p) at 12 MeV while Figs. \ref{fig-ci}(b) and (d) correspond to $^{48}$Ca(p,p) at 21 MeV. Expectedly, confidence intervals are narrower for the elastic angular distribution when elastic angular distribution data are used. The same principal is true when polarization data are used: confidence intervals are narrower for $iT_{11}$. 
However, these results alone do not provide sufficient basis to establish which data set contains more information and provide a better constraint on the optical model. These conclusions are also true for $^{48}Ca$(n,n) at 12 MeV (not shown).

\subsection{Sensitivity study with derivatives}
\label{sensitivity}

\begin{figure}[t]
\begin{center}
\includegraphics[width=0.49\textwidth]{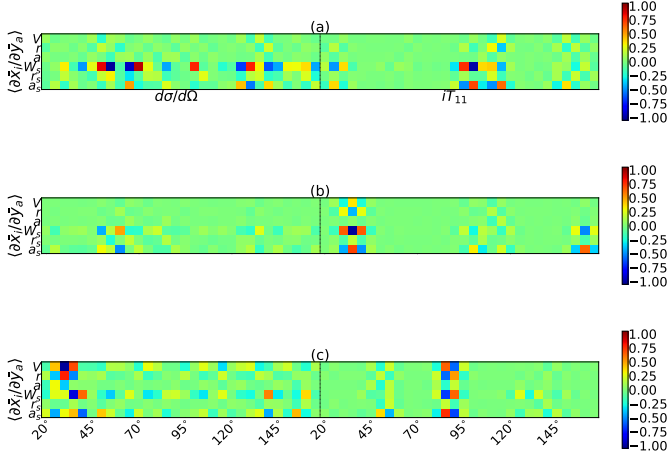}
\end{center}
\caption{Sensitivity matrix for $\frac{d\sigma}{d\Omega}$ data  and  $iT_{11}$  data: a) $^{48}$Ca(n,n) at 12 MeV; b) $^{48}$Ca(p,p) at 12 MeV and c) $^{48}$Ca(p,p) at 21 MeV. More details can be found in the text.}
\label{fig-sen}
\end{figure}

We next consider the sensitivities to understand which observables lead to the largest variation in the parameters. These sensitivities are drawn from the posteriors obtained using both the angular distribution and polarization data as constraints in the MCMC.
Fig. \ref{fig-sen} displays the sensitivities for $^{48}$Ca(n,n) at 12 MeV (panel a); $^{48}$Ca(p,p) at 12 MeV (panel b) and $^{48}$Ca(p,p) at 21 MeV (panel c). Comparing the sensitivity of the angular distributions (Fig. \ref{fig-sen} left) with the sensitivity of the polarization data (Fig. \ref{fig-sen} right), it is only for $^{48}$Ca(p,p) at 12 MeV that there is an indication that the polarization data offers a better constraint.

For the other cases the most notable feature is that the sensitivity occurs at different angles.
We can also compare the sensitivities for $^{48}$Ca(n,n) and $^{48}$Ca(p,p) at 12 MeV: our results suggest that neutrons offer a better constraint on the optical potential parameters at these energies, specifically for the imaginary term. At 12 MeV, the sensitivity of either observable to the real part is weak. At $21$ MeV, the sensitivity  of the parameters to $^{48}$Ca(p,p) increases significantly, particularly at forward angles for   $\frac{d\sigma}{d\Omega}$ and $\theta \approx 100$ degrees  for $iT_{11}$. At this higher energy, we find that both observables are  capable of constraining the parameters of the real part of the optical potential, in addition to the imaginary term. These quantitative results are consistent with the common understanding in the field.

Although we do not include details, we did perform a similar analysis for $^{208}$Pb(n,n) at 30 MeV, $^{208}$Pb(p,p) at 30 MeV and $^{208}$Pb(p,p) at 61 MeV. The results obtained for the sensitivities for these reactions also show variability with beam  energy: there are larger sensitivities to the elastic angular distributions for $^{208}$Pb(p,p) at 30 MeV (mostly on $W_s$ and $a_s$) compared to the polarization data. At $^{208}$Pb(p,p) at 61 MeV, the opposite is true.

\subsection{Bayesian evidence and Bayes factor}
\label{evidence}

Finally, we consider directly the Bayesian evidence to contrast the information content of cross section angular distribution data and polarization data within our model. 
\begin{table}[t]
\begin{center}
\begin{tabular}{|c|r|r|r|}
\hline Reaction & $\bar p(d\sigma/ d \Omega|\mathcal{M})$ & $\bar p(iT_{11}|\mathcal{M})$ & $ R$  \\ \hline
$^{48}$Ca(n,n) at 12 MeV & $0.833$ & $0.905$ & 1.09
  \\ \hline
$^{48}$Ca(p,p) at 12 MeV & $1.039$ & $1.208 $ & 1.16
  \\ \hline
$^{48}$Ca(p,p) at 21 MeV & $1.207$ & $0.602$ &0.50
  \\ \hline
$^{208}$Pb(n,n) at 30 MeV & $0.132$ & $0.052$ &0.39
  \\ \hline
$^{208}$Pb(p,p) at 30 MeV &$1.437$ & $0.403$ &0.28
  \\ \hline
$^{208}$Pb(p,p) at 61 MeV  &$0.051$ & $0.073 $ &1.44
  \\ \hline
\end{tabular}
\caption{Bayesian evidence averaged over angle (multiplied by $10^{-3}$) for the different reactions considered: using only cross section data (2nd column), using only polarization data (3rd column), and the ratio between the averaged Bayesian evidence using polarization data over that with cross section data (the Bayes' factor). 
%\IR{Updated table with results for (V,r,a,Ws,rs,as) to be consistent with the rest of the sections}
}
\label{tab-evidence}
\end{center}
\end{table}
$N=1,500,000$ is used to compute the evidence $p(d|\mathcal{M})$  for all cases. These are then averaged over angle to obtain $\bar p(d|\mathcal{M})$. The results are provided in Table \ref{tab-evidence}: the evidence obtained for cross section data (column 2) can be easily compared to that obtained for the polarization data (column 3). The Bayes' factor, defined as the ratio $R= \bar p(iT_{11}|\mathcal{M})/\bar p(d\sigma/ d \Omega|\mathcal{M})$ of the two average evidences, is shown in the last column.  Although we did not show the details of the calibrations for the reactions on Pb,  we include in the table the results for the evidence obtained for $^{208}$Pb(n,n) at 30 MeV, $^{208}$Pb(p,p) at 30 MeV and $^{208}$Pb(p,p) at 61 MeV using the same setup at was used for the $^{48}$Ca reactions.

For each reaction considered, the evidence for $d \sigma/ d \Omega$  data and for  $iT_{11}$ data are  of the same order of magnitude.
%with the exception of the proton elastic scattering on $^{48}$Ca at $12$ MeV. For this one specific case, the evidences differ by an order of magnitude, and therefore the Bayes factor  becomes worth mentioning \cite{wiki}. This larger Bayes factor corroborates the results for the sensitivities where we had  shown that optical model parameters were significantly more sensitive to the polarization data than to the angular distributions. } 
%\IR{ this would be the case for the table correponding to V,r,a, W,ri, ai. but we updated the table to be the results for V,r,a,Ws,rs,as!}
We can also compare evidences for the same reaction at different energies. Clearly, in the context of this optical model, the cross section distribution for $^{48}$Ca(p,p) at $12$ MeV has less information than $^{48}$Ca(p,p) at $21$ MeV and the cross section distribution for $^{208}$Pb(p,p) at $30$ MeV has more information than $^{208}$Pb(p,p) at $61$ MeV. Although beyond the scope of this work, a study including a wider range of target nuclei and beam energies is necessary to understand systematic trends.

We should highlight that the evidence is essentially a very different measure than the $\frac{\partial \bar{x}_i}{\partial \bar{y}_a}$. The sensitivities in this work were obtained from averages over the posterior distributions and therefore show the rate of change in the region of parameter space defined by the posterior distributions. The evidence is an integral over the full parameter space, weighted by the likelihood. Longer tails in the likelihood result in larger evidence. Larger information content as measured in the model evidence, does not  necessarily translate to tighter constraints on the parameters themselves. Both tools should be used in concert.

%%%%%%%%%%%%%%%%%%%%%%%%%%%%%%%%%%%%%%%%%%%%%%%%%%%%%%%%%%

\subsection{Comparing real data and mock data}
\label{realdata}

\begin{figure}[t]
\begin{center}
\includegraphics[width=0.45\textwidth]{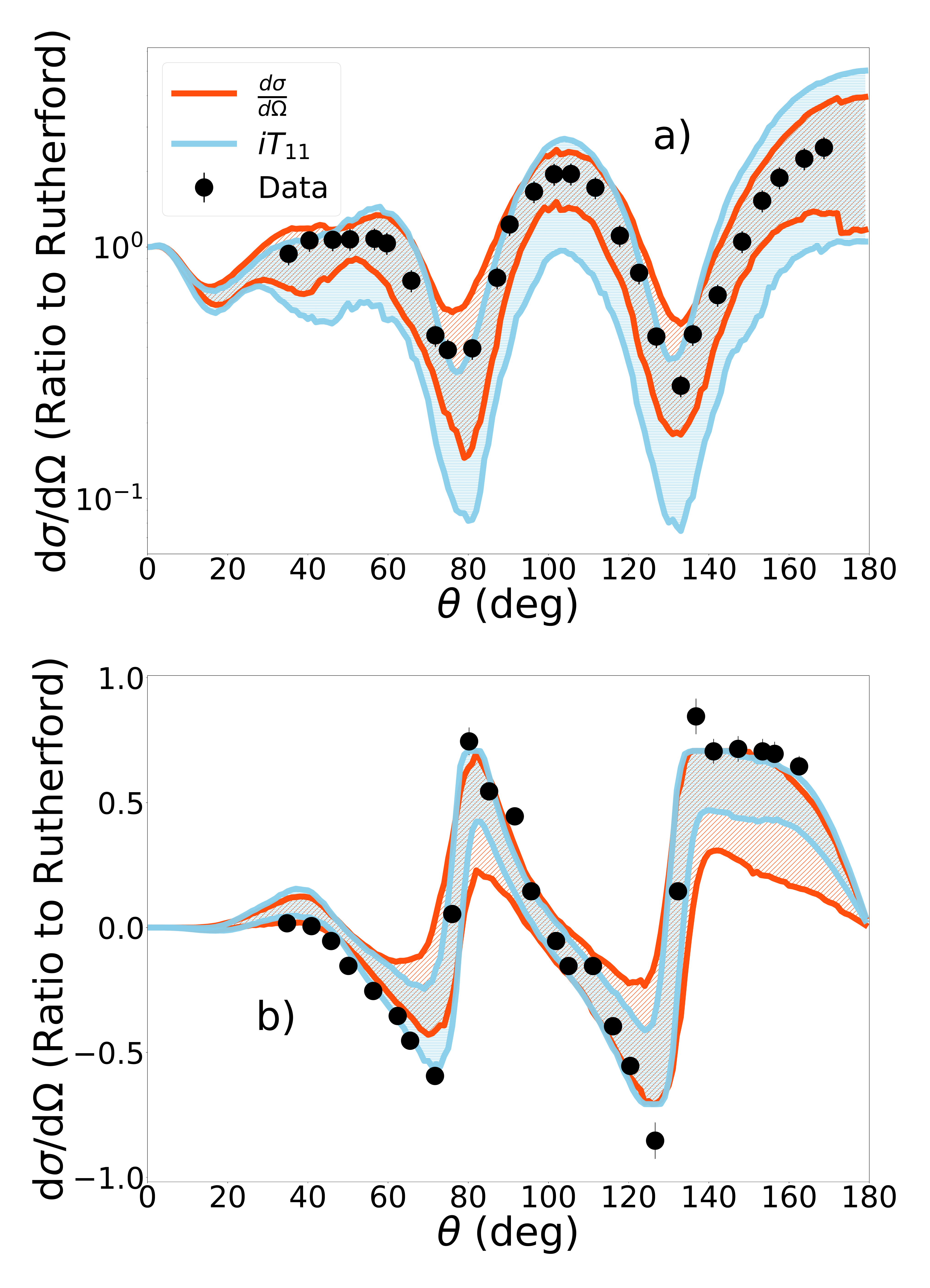}
\end{center}
\caption{$^{48}$Ca(p,p) at 14 MeV 95\% confidence intervals calculated using $\frac{d\sigma}{d\Omega}$ data (orange bands)  or  $iT_{11}$  data (blue bands) in the optimization procedure: a) $d\sigma/d\Omega$ and b) $iT_{11}$. Data from \cite{realdata}.}
\label{fig-realdata}
\end{figure}

All results presented so far involve mock data. The choice for mock data in this work is based on the control it provides: we can produce data at any energy, simultaneously have elastic angular distributions and polarization data across the whole angular range. However one might be concerned that results with mock data do not represent the real world. It is understood that the KD parameterization cannot exactly reproduce elastic scattering data for a given case. Our point here is that KD is close enough to reality to provide a good illustration for these new statistical tools.

We therefore pick an example to demonstrate that real data and mock data can lead to qualitatively similar results. For more detail in the comparison of mock and real data see \cite{lovell2020}.
We found data close to $^{48}$Ca(p,p) at 12 MeV, corresponding to a reaction with protons at $14$ MeV \cite{realdata}.  In Fig. \ref{fig-realdata} we show the angular distribution of the cross section and the polarization for protons on $^{48}$Ca at 14 MeV and verify these are very similar to those shown in Fig.\ref{fig-ci} for the corresponding reaction at $12$ MeV. As before, the orange bands (blue bands) correspond to the 95\% confidence intervals when the elastic cross section data (polarization data) are used in the fit. As in the case for mock data, here we also find that the uncertainty in the elastic angular distribution is smaller when $d\sigma/d\Omega$ data is used. Conversely, the uncertainty in the $iT_{11}$ distribution is smaller when polarization data is used.

It should not be understood from this comparison that mock data can replace real data.  Although qualitatively similar,  there are significant quantitative differences in the posterior distributions obtained with mock data compared to real data. Thus, in a practical application of these tools, one should always use real data in the statistical analysis.

%%%%%%%%%%%%%%%%%%%%%%%%%%%%%%%%%%%%%%%%%%%%%%%%%%%%%%%%%%%%%%%%%

\section{Conclusions}
\label{conclusions}

In this work we have explored three statistical tools in the context of nuclear scattering that allow us to go beyond uncertainty quantification toward  understanding sensitivities of the parameters and information content of reaction observables. 
We consider the principal component analysis in observable space, the parameter to observable sensitivities and  the Bayesian evidence. 
To introduce these tools, we construct two limiting toy models, an optical model just with surface absorption and an optical model just with volume absorption. We perform standard MCMC calculations varying six optical model parameters, and constrain each model with the angular distributions for elastic scattering. We obtain the Bayesian parameter posteriors  distributions and the associated confidence intervals for the angular distributions. We then apply the statistical tools and find that both sensitivities and Bayesian evidence provide important insights in  discriminating between models. 

Next we repeat this process for realistic cases and using either  $\frac{d\sigma}{d\Omega}$ or  $iT_{11}$ to constrain the optical parameters.
We study $^{48}$Ca(n,n) at 12 MeV, $^{48}$Ca(p,p) at 12 MeV and $^{48}$Ca(p,p) at 21 MeV.   Neither the confidence intervals nor the parameter posterior distributions help in determining which observable is best to constrain the  optical model parameters. We also did not find the principal component analysis defined in terms of angles to be useful, since it produced roughly equal weights  for all components. 

In contrast,  sensitivities provided important insights. 
For most examples studied, $\frac{d\sigma}{d\Omega}$ and $iT_{11}$ are sensitive to the same parameters, and to the same degree. However they provide constraints in different angular regions. The exception being the reaction $^{48}$Ca(p,p) at 12 MeV, for which the differential cross sections offer less sensitivity than the polarization data.

Finally, we also computed the Bayesian evidence for each reaction. The integral over the parameter space had to be performed fully numerically, as the assumptions of Gaussian distributions for the analytic approximations were not valid. We found that in order to get converged values for the evidence, a much larger number of draws was necessary as compared to the statistics collected for the parameter posteriors and confidence intervals. We compared the values of the evidence, averaged over angle, obtained when the cross section angular distribution was used as a constraint with those when using the polarization data. 
While Bayes' factor (the ratio of the evidence using $iT_{11}$ over that using $\frac{d\sigma}{d\Omega}$) shows some variability, for most reactions studied it is of order one, and therefore not significant. From our results, we conclude that, within our optical model, the information content of the differential cross sections and the polarization data is roughly the same. 

The data we included ($d \sigma/ d \Omega$ and $iT_{11}$) are both associated with the same elastic scattering channel. Future plans include the application of these tools to situations where we scrutinize between data that are more dissimilar, such as  elastic scattering and charge-exchange, knockout or breakup.
One essential ingredient for these advances is the speed-up in computations. In this regard, recent work on emulators (e.g. \cite{furnstahl2020,drischler2021}) holds much promise.

\vspace{1cm}

\begin{acknowledgments}
The authors would like to thank S.~Pratt and M. Hirn for  useful discussions on the sensitivity analysis and PCA, and L.~Neufcourt for discussions on the Bayesian evidence.
This work was supported by the U.S. Department of Energy (Office of Science, Nuclear Physics) under grant DE-SC0021422, the National Science Foundation under Grant  PHY-1811815, and was performed under the auspice of the U.S. Department of Energy by Los Alamos National Laboratory under Contract 89233218CNA000001.  This work relied on iCER and the High Performance Computing Center at Michigan State University for computational resources. 
\end{acknowledgments}
%%%%%%%%%%%%%%%%%%%%%%%%%%%%%%%%%%%%%%%%%%%%%%%%%%%%%%%%%%%%%%%%%%%%%%%%%%%%%%%%%%%%%%%%%%%%%%%%%

\bibliography{uq} 

\end{document}